\documentclass[aps,prl,reprint,showpacs,superscriptaddress]{revtex4-1}

\usepackage{tikz}
\usepackage{amsmath}
\usepackage{hyperref}
\usepackage{cleveref}
\usepackage{upgreek}
\hypersetup{
    colorlinks,
    citecolor=blue,
    linkcolor=blue,     urlcolor=blue
}

\renewcommand{\vec}[1]{\boldsymbol{#1}}
\newcommand{\vgal}{\vec{v_{\textrm{gal}}}}
\newcommand{\nab}{\vec{\nabla'}}
\newcommand{\Dt}[1]{ \frac{\partial #1}{\partial t}}
\newcommand{\mc}[1]{\hat{\mathcal{#1}}}

\begin{document}


\title{Stable discrete representation of relativistically drifting plasmas}


\author{M. Kirchen}
\email{manuel.kirchen@desy.de}
\affiliation{Center for Free-Electron Laser Science \& Department of Physics,
University of Hamburg, 22761 Hamburg, Germany}
\author{R. Lehe}
\affiliation{Lawrence Berkeley National Laboratory, Berkeley, CA 94720, USA}
\author{B.\,B. Godfrey}
\affiliation{Lawrence Berkeley National Laboratory, Berkeley, CA 94720, USA}
\affiliation{University of Maryland, College Park, MD 20742, USA}
\author{I. Dornmair}
\affiliation{Center for Free-Electron Laser Science \& Department of Physics,
University of Hamburg, 22761 Hamburg, Germany}
\author{S. Jalas}
\affiliation{Center for Free-Electron Laser Science \& Department of Physics,
University of Hamburg, 22761 Hamburg, Germany}
\author{K. Peters}
\affiliation{Center for Free-Electron Laser Science \& Department of Physics,
University of Hamburg, 22761 Hamburg, Germany}
\author{J.-L. Vay}
\affiliation{Lawrence Berkeley National Laboratory, Berkeley, CA 94720, USA}
\author{A.\,R. Maier}
\affiliation{Center for Free-Electron Laser Science \& Department of Physics,
University of Hamburg, 22761 Hamburg, Germany}



\date{\today}

\begin{abstract}
Representing the electrodynamics of relativistically drifting particle ensembles in discrete, co-propagating \emph{Galilean} coordinates enables the derivation of a Particle-in-Cell algorithm that is intrinsically free of the Numerical Cherenkov Instability for plasmas flowing at a uniform velocity. Application of the method is shown by modeling plasma accelerators in a Lorentz-transformed optimal frame of reference.
\end{abstract}


\maketitle
Describing complex physics beyond analytical theories requires numerical modeling of the underlying equations in discrete space. In plasma physics, astrophysics or accelerator physics, Particle-In-Cell (PIC) methods are commonly used to self-consistently solve the electromagnetic interaction of particle ensembles \cite{BunemanJCP1980, DawsonRMP1983, Hockney1988, Birdsall2004}. A PIC algorithm iteratively solves Maxwell's equations on a discrete grid with particles following the equations of motion in a continuous space.

Some of the physical systems accessible with the PIC method feature plasmas drifting at relativistic velocities, for example when modeling plasma-based particle accelerators \cite{EsareyRMP2009} in the optimal frame of reference \cite{VayPRL2007} or astrophysical plasma interactions \cite{SironiAJ2013}. In those cases, the applicability of the to-date electromagnetic PIC algorithms is fundamentally limited by the Numerical Cherenkov Instability (NCI) \cite{GodfreyJCP1974,GodfreyJCP1975,GodfreyJCP2013,XuCPC2013}, which either falsifies the numerical results or causes virulent growth of unphysical waves. 

Here, we present a novel discrete formulation of the fundamental kinetic equations of plasmas, i.e.\ Maxwell's and the Newton-Lorentz equations, that represents the physics in a moving \textit{Galilean} frame of reference and thereby is intrinsically free of the NCI for plasmas drifting at uniform relativistic velocities. 

The NCI originates from the coupling of distorted electromagnetic modes with spurious particle modes. Distortions of the electromagnetic field modes are caused by numerical inaccuracies of the discretized field-solving algorithm. Spurious spatial and temporal aliases of the physical particle modes result from the numerical mismatch of sampling the continuously distributed particle quantities to the discrete field grid. To first order, for example, Numerical Cherenkov Radiation (NCR) can occur, if the dispersion relation of the electromagnetic waves is numerically distorted. In this case, particles moving at relativistic velocities $v_p \approx c$ couple resonantly to electromagnetic waves of high frequency, which propagate at a spurious phase velocity $v_\Phi < v_p < c$, causing Cherenkov-like radiation to be emitted. Although many algorithms, such as pseudo-spectral solvers \cite{Haber}, do not suffer from NCR, higher order NCI effects severely limit the stable modeling of relativistic plasmas.

So far, no electromagnetic, fully explicit PIC algorithm is intrinsically free of NCI, even for the simple case of a plasma drifting at a uniform relativistic velocity. Previously developed suppression strategies can limit the NCI growth rate, thereby retaining the physical meaningfulness of a simulation. For example, wide-band smoothing \cite{VayAAC10,VayJCP2011,VayPOPL11} or damping \cite{MartinsCPC10} of the currents or electromagnetic fields can hinder the development of the instability. Coupling of unphysical modes can be mitigated by slightly changing the ratio of the electric and magnetic fields as seen by the particles \cite{GodfreyJCP2014b,Godfreyarxiv2014,GodfreyCPC2015}, by scaling the deposited currents with a frequency-dependent factor \cite{GodfreyJCP2014,GodfreyIEEE2014}, or by artificially modifying the physical electromagnetic dispersion relation \cite{YuCPC2015,YuCPC2015-Circ,Yu-arxiv2016}. Yet, all of these techniques rely on numerical methods that potentially alter the physics and could affect the results obtained with the algorithm.

In contrast, the method presented in this paper inherently eliminates the NCI for a relativistically drifting plasma, as opposed to suppressing its growth by the measures described above. From a heuristic point of view, the main difference between modeling a plasma at rest, showing no NCI, and a relativistically drifting plasma, is that the particles move with respect to the static numerical grid. Thus, intuitively, by mathematically representing the underlying discrete equations such that this discrepancy in relative movement is eliminated, the NCI should be suppressed.

This is achieved by applying a Galilean coordinate transformation of the form
\begin{equation}
\vec{x}' = \vec{x} - \vgal t \nonumber
\end{equation}
to the frame of reference in which a plasma is moving at a relativistic velocity. Consequently, the equations of motion and Maxwell's equations transform to

\begin{align}
\frac{d\vec{x}'}{dt} &= \frac{\vec{p}}{\gamma m} - \vgal, \nonumber \\
\frac{d\vec{p}}{dt} &= q \left( \vec{E} +
\frac{\vec{p}}{\gamma m} \times \vec{B} \right), \nonumber \\
\left( \Dt{\;} - \vgal\cdot\nab\right)\vec{B} &= -\nab\times\vec{E}, \nonumber \\
\frac{1}{c^2}\left( \Dt{\;} - \vgal\cdot\nab\right)\vec{E} &= \nab\times\vec{B} - \mu_0\vec{j}, \nonumber 
\end{align}
and the continuity equation becomes $\left( \partial_t - \vgal\cdot\nab \right)\rho + \nab\cdot \vec{j} = 0$. Here, $\nab$ denotes the spatial derivative with respect to the \emph{Galilean} coordinates $\vec{x}'$. For $\vec{v}_\textrm{gal} = \vec{0}$, these equations reduce to their well-known original form.

Using the Pseudo-Spectral Analytical Time Domain (PSATD) framework \cite{Haber}, the last two equations are transformed to Fourier space and can then be integrated \emph{analytically} in time. As the quantities are only known at discrete times in a PIC algorithm, the time evolution of $\rho$ and $\vec{j}$ needs to be explicitly taken into account during integration. Typically, the currents are assumed to be constant over one time step $\Delta t$ in the original coordinates $\vec{x}$. A key difference of our new scheme is that we assume the currents to be co-moving with respect to the original coordinates $\vec{x}$, hence constant over one time step in the Galilean coordinates $\vec{x}'$. The resulting Galilean-PSATD equations for the advance of the spectral field components, $\vec{\mc{E}}$ and $\vec{\mc{B}}$, from time step $n \Delta t$ to $(n+1) \Delta t$ are then given by (see \cite{Lehe2016} for a derivation)

\begin{align}
\vec{\mc{B}}^{n+1} &= \theta^2 C \vec{\mc{B}}^n
 -\frac{\theta^2 S}{ck}i\vec{k}\times \vec{\mc{E}}^n
+ \;\frac{\theta \chi_1}{\epsilon_0c^2k^2}\;i\vec{k} \times
                     \vec{\mc{J}}^{n+1/2}, \nonumber \\
\vec{\mc{E}}^{n+1} &=  \theta^2 C  \vec{\mc{E}}^n
 +\frac{\theta^2 S}{k} \,c i\vec{k}\times \vec{\mc{B}}^n 
+\frac{i\nu \theta \chi_1 - \theta^2S}{\epsilon_0 ck} \; \vec{\mc{J}}^{n+1/2} \nonumber \\[-5px]
& - \frac{1}{\epsilon_0k^2}\left(\; \chi_2\;\mc{\rho}^{n+1} -
  \theta^2\chi_3\;\mc{\rho}^{n} \;\right) i\vec{k}, \nonumber \\
C &= \cos(ck\Delta t), \quad S = \sin(ck\Delta t), \quad k
= |\vec{k}|, \nonumber \\
\nu &= \frac{\vec{k}\cdot\vgal}{ck}, \quad \theta =
  e^{i\vec{k}\cdot\vgal\Delta t/2}, \quad \theta^* =
  e^{-i\vec{k}\cdot\vgal\Delta t/2}, \nonumber \\
\chi_1 &=  \frac{1}{1 -\nu^2} \left( \theta^* -  C \theta + i
  \nu \theta S \right), \nonumber \\
\chi_2 &= \frac{\chi_1 - \theta(1-C)}{\theta^*-\theta}, \quad
\chi_3 = \frac{\chi_1-\theta^*(1-C)}{\theta^*-\theta}, \nonumber
\end{align}
where $\vec{k}$ is the wavevector. The currents $\vec{\mc{J}}$ at time $(n+1/2) \Delta t$ and the charge density $\mc{\rho}$ at time $n \Delta t$ and $(n+1) \Delta t$ are generated by the particles and deposited to the grid nodes before being transformed to Fourier space. Subsequently, the updated fields are transformed back to real space and interpolated to the particles, which are then advanced in time using the Galilean transformed equations of motion.

\begin{figure}[tbh]
\includegraphics[width=0.9\columnwidth]{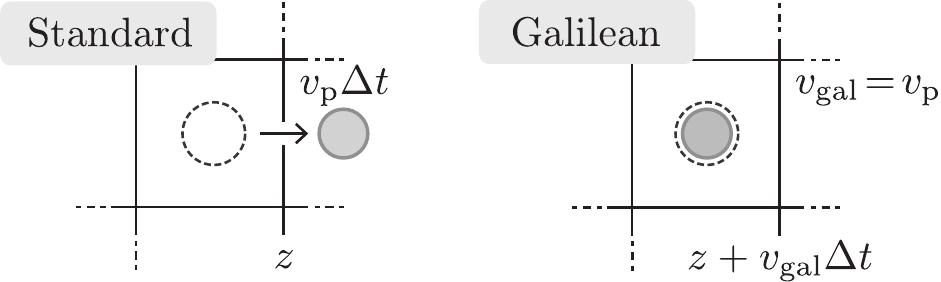}
\caption{Schematic drawing illustrating the \emph{Galilean} concept. Without applying a Galilean coordinate transformation to the Particle-In-Cell equations (\emph{Standard}), a plasma flowing with velocity $v_\textrm{p}$ in $z$ (represented by a single particle) would propagate a distance $v_\textrm{p} \Delta t$ with respect to the numerical grid (represented by a single cell) during one time step $\Delta t$. However, in a Galilean transformed discrete space $\vec{x}'$ with $\vgal = (0,\,0,\,v_\textrm{gal}=v_\textrm{p})$ the plasma particles remain static with respect to the discrete grid nodes, which themselves propagate a distance $z+v_\textrm{gal} \Delta t$ in the original coordinate system $\vec{x}$.} \label{fig:fig1}
\end{figure}

This algorithm allows to model a plasma moving at $\vec{v}_{p}$ in a co-propagating set of coordinates $\vec{x}'$ with $\vec{v}_\textrm{gal} = \vec{v}_{p}$. As shown in fig.\ \ref{fig:fig1}, the flowing plasma particles now remain static with respect to the numerical grid. Because of this and the \emph{co-moving current assumption}, the NCI is completely eliminated for particles streaming at the velocity $\vec{v}_{p}$.

The algorithm is implemented in the \textsc{Warp} code \cite{Warpref}, for Cartesian coordinates, as well as in the recently developed quasi-cylindrical \cite{Lifschitz} code \textsc{Fbpic} \cite{LeheCPC2016}. In \cite{Lehe2016} we also present an analytical derivation of the dispersion relation and conduct a detailed empirical and theoretical stability analysis for uniform relativistically flowing plasmas. Here, we restrict ourselves to presenting the general concept and the practical demonstration of the stability and accuracy of our new method with a direct application. In the following, Lorentz-boosted frame simulations of plasma acceleration with \textsc{Fbpic} are presented.

Plasma-based accelerators \cite{EsareyRMP2009} can sustain high field gradients, allowing for the acceleration of charged particles within distances shorter by orders of magnitude compared to conventional accelerators. In a plasma-wakefield accelerator, an intense driver beam (a high intensity laser pulse or particle bunch) propagates through an underdense plasma and induces a charge separation on the sub-mm scale. This leads to the excitation of a trailing density wave carrying large electric fields, suitable for the acceleration of electron bunches to high energies.

The natural frame of reference for PIC simulations of plasma accelerators is the laboratory frame. In this frame of reference the physical objects of small scale, i.e.\ the laser or particle beam, propagate at relativistic velocities in a single direction while interacting with a large scale object that is static, i.e.\ the plasma. A Lorentz transformation in the propagation direction of the driver beam then relaxes the requirements on the spatial resolution while contracting the required simulation distance \cite{VayPRL2007}. In this Lorentz-boosted frame, the co-propagating quantities, e.g.\ the laser or the plasma wavelength, are elongated by $\gamma(1+\beta)$, whereas the previously static lengths, such as the plasma, are contracted by $\gamma$ and counter-propagate with a relativistic velocity $-\beta c$. Thereby, a speed-up by orders of magnitude can be achieved that scales as $\propto \gamma_{\textrm{boost}}^2$, with the maximum speed-up typically limited to $\approx 2 \gamma_{\textrm{wake}}^2$, i.e.\ the phase velocity of the wake, in the case of laser-plasma acceleration.

In the following, we show simulations of a non-linear plasma wave driven by a laser pulse with wavelength $\lambda=800\,$nm, peak normalized vector potential $a_0=1.5$, pulse length $c\tau=8\,\upmu$m and waist $w_0=30\,\upmu$m that propagates through a matched plasma guiding channel with an on-axis electron density $n_e=1\cdot10^{18}\,\textrm{cm}^{-3}$. In the generated wakefields, a $1\,$pC electron bunch of size $\sigma_z=1\,\upmu$m, $\sigma_r=2\,\upmu$m, and normalized emittance $\epsilon_\textrm{n}=0.5\,$mm$\,$mrad, located at the back of the first wave bucket, is accelerated from $100\,$MeV to $687\,$MeV within a propagation distance of $\textrm{z}_{\textrm{prop}} \approx 14.3\,$mm. The resolution of the simulation is $40$ cells per $\upmu$m in the longitudinal and $2$ cells per $\upmu$m in the transverse direction. Third order particle shapes are used with 24 particles per cell. The time step is set to $\Delta t = \Delta z/c$.

As described above, the occurrence of the NCI, caused by the counter-streaming relativistic plasma, can hinder the application of the Lorentz-boosted frame method for simulations of plasma-wakefield accelerators. With our new method, however, such a simulation is modeled in a Galilean transformed coordinate system that counter-propagates to the Lorentz-boosted frame with the velocity $v_\textrm{gal} = -\beta c$ in the direction of the boosted plasma. With respect to the numerical grid, the background plasma is thus static, whereas the elongated quantities, such as the laser pulse and the electron bunch, propagate with a velocity increased by the same amount with respect to the grid.

\begin{figure}[tbh]
\includegraphics[width=1.0\columnwidth]{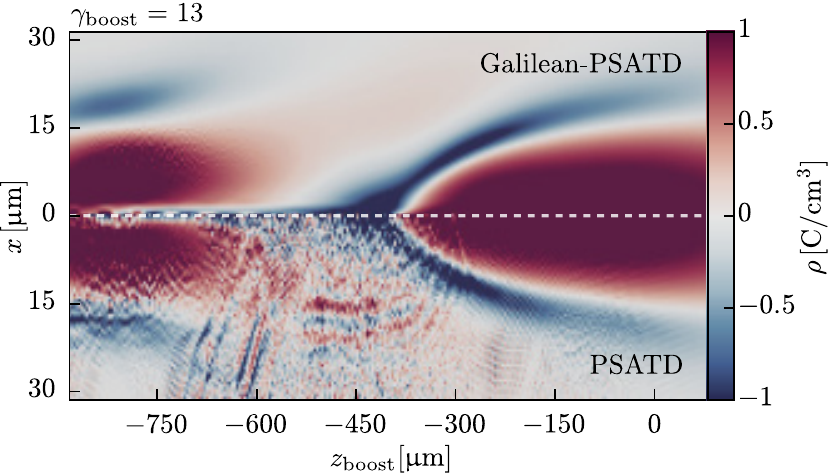}
\caption{Charge density $\rho$ obtained from a Lorentz-boosted frame simulation ($\gamma_{\textrm{boost}} = 13$) of a non-linear laser-plasma wave. At the time step shown, a part of the laser pulse, propagating to the right, has already left the plasma, which is flowing to the left. The upper-half corresponds to a Galilean-PSATD simulation with $v_\textrm{gal} = -\beta c$, showing no instability. The lower half shows the same simulation, mirrored along the $x=0$ axis, but conducted with the standard PSATD solver. Here, a fast growing, virulent NCI can be observed.}
\label{fig:fig2}
\end{figure}
Fig.\ \ref{fig:fig2} shows the charge density obtained in a Lorentz boosted frame ($\gamma_{\textrm{boost}} = 13$) with boosted longitudinal coordinate $z_\textrm{boost}=\gamma(z-vt)$. The upper-half of the plot shows the results of a simulation with the Galilean-PSATD solver, whereas the lower-half shows the corresponding results of a standard PSATD simulation. Here, a fast growing NCI can be observed. In contrast, the same simulation remains completely stable when modeled in the Galilean transformed discrete space. We emphasize that all numerical parameters are the same in these simulations, except for the difference in using $v_\textrm{gal} = -\beta c$ instead of $v_\textrm{gal}=0$ for the Galilean-PSATD equations. Thus, the absence of the instability results solely from the Galilean transformation of the underlying discrete equations. Even though the electron bunch and the grid move in opposite directions, we do not observe any NCI around the bunch. This can be explained by the fact that the electron bunch has a density that is much lower than the plasma, as it is elongated in the Lorentz-boosted frame. Moreover, due to its non-zero charge, it is probably much less affected by higher order Numerical Cherenkov effects. Likewise, in a laboratory frame simulation, a relativistic electron bunch does typically not lead to an instability, as long as NCR is suppressed \cite{LehePRSTAB2013}.

\begin{figure}[tbh]
\includegraphics[width=1.0\columnwidth]{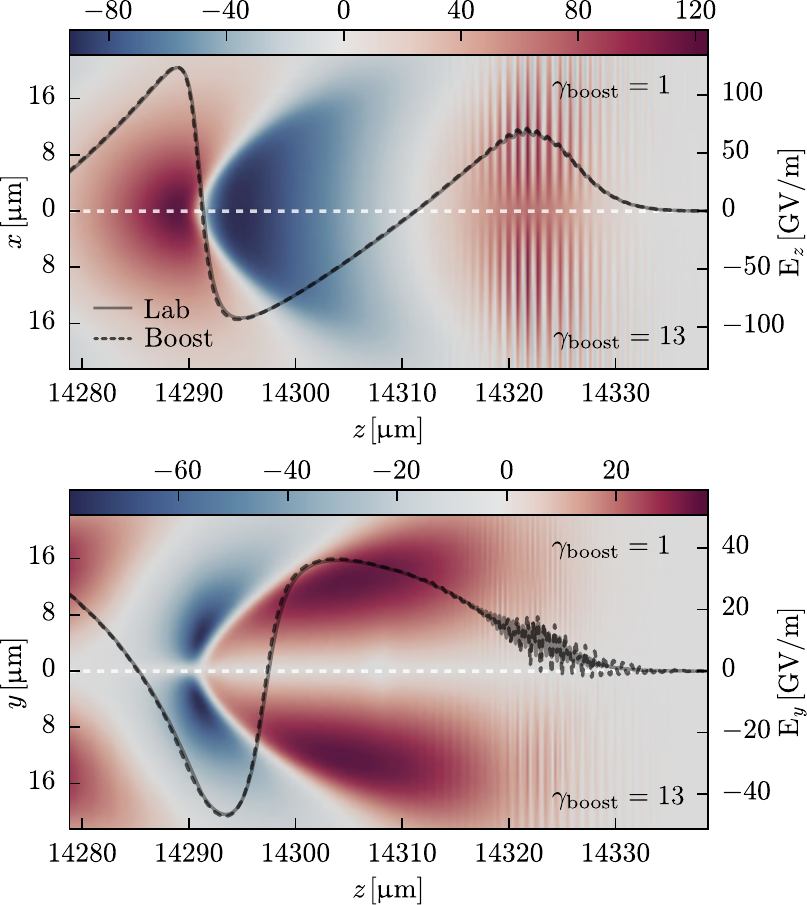}
\caption{Comparison of the accelerating fields ($E_z$ fields and on-axis lineout) and the focusing fields ($E_y$ fields and off-axis, $y = 11.25\,\upmu$m, lineout). The upper-half of the plots shows the results of a laboratory frame simulation (solid line) with $\gamma_\textrm{boost}=1$. The lower-half shows the back-transformed results of a Lorentz-boosted frame simulation (dashed line) with $\gamma_\textrm{boost} = 13$, mirrored along the $x,y=0$ axis.}
\label{fig:fig3}
\end{figure}
In order to validate the accuracy of our new method, results from the stable Lorentz-boosted frame simulation are compared to a laboratory frame simulation. Fig.\ \ref{fig:fig3} shows the electric fields at the end of the acceleration distance. The upper-half of the plots shows the results of the reference simulation ($\gamma_\textrm{boost}=1$), whereas the lower-half shows the corresponding back-transformed results of the Lorentz-boosted frame simulation ($\gamma_\textrm{boost}=13$). Both the longitudinal fields $E_z$ and the transverse fields $E_y$ show no differences. Note that the results in the Lorentz-boosted frame are obtained within only a few thousand time steps, whereas the lab frame simulation takes more than half a million time steps to complete. We achieve a speed-up of $\approx\,$287 ($\approx\,$92$ \%$ of the optimal speed-up) with \textsc{Fbpic}, where the only overhead is the on-the-fly back-transformation of data to the laboratory frame.

\begin{figure}[tbh]
\includegraphics[width=1.0\columnwidth]{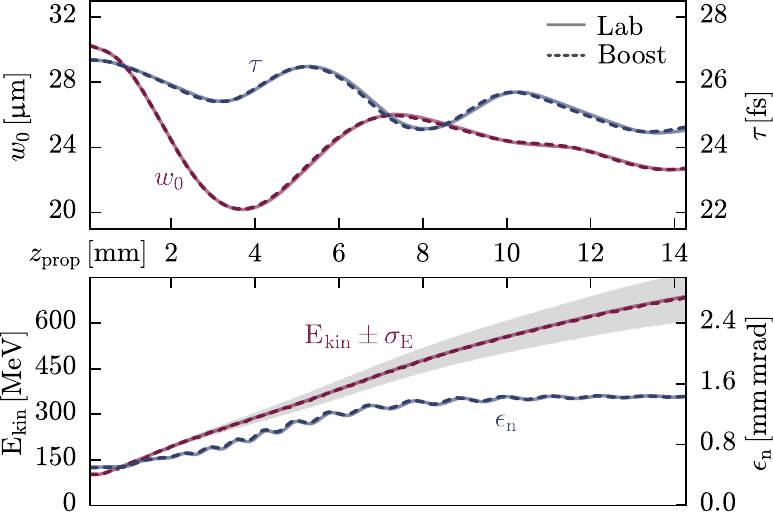}
\caption{Comparison of the laser and electron bunch evolution between the laboratory frame (solid line) and the Lorentz-boosted frame (dashed line) simulation. The upper plot shows the pulse duration $\tau$ (blue) and the laser waist $w_0$ (red) and the lower plot shows the kinetic Energy $\textrm{E}_\textrm{kin}$ (red), the rms energy spread $\sigma_\textrm{E}$ (gray area) and the normalized emittance $\epsilon_\textrm{n}$ (blue) over the complete acceleration distance of $z_\textrm{prop} \approx 14.3\,$mm.}
\label{fig:fig4}
\end{figure}
Furthermore, we compare characteristic bunch and laser parameters to demonstrate that the physics is preserved in the Lorentz-boosted frame. Fig.\ \ref{fig:fig4} shows the evolution of the laser waist $w_0$ and the pulse duration $\tau$, as well as the kinetic energy $\textrm{E}_\textrm{kin}$, the rms energy spread $\sigma_\textrm{E}$ and the normalized emittance $\epsilon_\textrm{n}$ of the accelerated electron bunch. During the propagation through the plasma guiding channel, the laser pulse self-focuses transversely and the pulse duration shortens due to the relativistic interaction with the plasma. The electron bunch is initially situated at the minimum of the accelerating field and slips towards the laser pulse during the propagation. It is accelerated to 687$\,$MeV, while accumulating an rms energy spread of $\approx\,$11.5$\,\%$, due to the slope of the accelerating field. As the bunch enters the plasma, strong transverse fields act on it abruptly, causing transverse oscillations of the bunch size and growth of the emittance $\epsilon_\textrm{n}$ to around $1.4\,$mm$\,$mrad. In direct comparison with the laboratory simulation, all the quantities shown differ only on the sub-percent level at the end of the propagation distance, which resembles a remarkable precision.

In conclusion, we have proposed a novel discrete formulation of the fundamental kinetic equations of plasmas in Galilean transformed coordinates. To the best of our knowledge, we thereby derived the first electromagnetic, fully explicit PIC representation that is intrinsically free of the NCI for plasmas flowing at a uniform velocity. Our concept is not reliant on otherwise inevitable numerical corrections and, unlike most of the previous NCI suppression strategies, it is independent of the specific geometry. This allows to combine the accuracy and efficiency of a spectral, quasi-3D PIC algorithm with the superior stability properties of the presented method. Applying the Galilean scheme to simulations of plasma accelerators in the Lorentz-boosted frame yields excellent agreement, while achieving a close-to-optimal speed-up of more than two orders of magnitude in practice.

Future research will cover the applicability of the Galilean scheme to other solvers, the parallelization based on domain decomposition \cite{VayJCP2013} with arbitrary-order spectral methods \cite{Birdsall2004, Vincenti2015, Yu-arxiv2016} and the potential generalization to support arbitrary relativistic plasma flows. For example, the new method could directly be extended to model collisionless astrophysical shocks \cite{SironiAJ2013} involving two plasmas, by employing separate numerical grids for each plasma using different Galilean transformed coordinates. Taking advantage of the superposition principle, only the electromagnetic fields would be shared between those individual grids.

\begin{acknowledgments}
We gratefully acknowledge the computing time provided on the supercomputer JURECA under project HHH20 and on the PHYSnet cluster of the University of Hamburg. Work at LBNL was funded by the Director, Office of Science, Office of High Energy Physics, U.S.\ Dept.\ of Energy under Contract No. DE-AC02-05CH11231, including from the Laboratory Directed Research and Development (LDRD) funding from Berkeley Lab.
\end{acknowledgments}

%


\begin{thebibliography}{31}%
\makeatletter
\providecommand \@ifxundefined [1]{%
 \@ifx{#1\undefined}
}%
\providecommand \@ifnum [1]{%
 \ifnum #1\expandafter \@firstoftwo
 \else \expandafter \@secondoftwo
 \fi
}%
\providecommand \@ifx [1]{%
 \ifx #1\expandafter \@firstoftwo
 \else \expandafter \@secondoftwo
 \fi
}%
\providecommand \natexlab [1]{#1}%
\providecommand \enquote  [1]{``#1''}%
\providecommand \bibnamefont  [1]{#1}%
\providecommand \bibfnamefont [1]{#1}%
\providecommand \citenamefont [1]{#1}%
\providecommand \href@noop [0]{\@secondoftwo}%
\providecommand \href [0]{\begingroup \@sanitize@url \@href}%
\providecommand \@href[1]{\@@startlink{#1}\@@href}%
\providecommand \@@href[1]{\endgroup#1\@@endlink}%
\providecommand \@sanitize@url [0]{\catcode `\\12\catcode `\$12\catcode
  `\&12\catcode `\#12\catcode `\^12\catcode `\_12\catcode `\%12\relax}%
\providecommand \@@startlink[1]{}%
\providecommand \@@endlink[0]{}%
\providecommand \url  [0]{\begingroup\@sanitize@url \@url }%
\providecommand \@url [1]{\endgroup\@href {#1}{\urlprefix }}%
\providecommand \urlprefix  [0]{URL }%
\providecommand \Eprint [0]{\href }%
\providecommand \doibase [0]{http://dx.doi.org/}%
\providecommand \selectlanguage [0]{\@gobble}%
\providecommand \bibinfo  [0]{\@secondoftwo}%
\providecommand \bibfield  [0]{\@secondoftwo}%
\providecommand \translation [1]{[#1]}%
\providecommand \BibitemOpen [0]{}%
\providecommand \bibitemStop [0]{}%
\providecommand \bibitemNoStop [0]{.\EOS\space}%
\providecommand \EOS [0]{\spacefactor3000\relax}%
\providecommand \BibitemShut  [1]{\csname bibitem#1\endcsname}%
\let\auto@bib@innerbib\@empty
\bibitem [{\citenamefont {Buneman}\ \emph {et~al.}(1980)\citenamefont
  {Buneman}, \citenamefont {Barnes}, \citenamefont {Green},\ and\ \citenamefont
  {Nielsen}}]{BunemanJCP1980}%
  \BibitemOpen
  \bibfield  {author} {\bibinfo {author} {\bibfnamefont {O.}~\bibnamefont
  {Buneman}}, \bibinfo {author} {\bibfnamefont {C.}~\bibnamefont {Barnes}},
  \bibinfo {author} {\bibfnamefont {J.}~\bibnamefont {Green}}, \ and\ \bibinfo
  {author} {\bibfnamefont {D.}~\bibnamefont {Nielsen}},\ }\href {\doibase
  10.1016/0021-9991(80)90010-8} {\bibfield  {journal} {\bibinfo  {journal}
  {Journal of Computational Physics}\ }\textbf {\bibinfo {volume} {38}},\
  \bibinfo {pages} {1 } (\bibinfo {year} {1980})}\BibitemShut {NoStop}%
\bibitem [{\citenamefont {Dawson}(1983)}]{DawsonRMP1983}%
  \BibitemOpen
  \bibfield  {author} {\bibinfo {author} {\bibfnamefont {J.~M.}\ \bibnamefont
  {Dawson}},\ }\href {\doibase 10.1103/RevModPhys.55.403} {\bibfield  {journal}
  {\bibinfo  {journal} {Rev. Mod. Phys.}\ }\textbf {\bibinfo {volume} {55}},\
  \bibinfo {pages} {403} (\bibinfo {year} {1983})}\BibitemShut {NoStop}%
\bibitem [{\citenamefont {Hockney}\ and\ \citenamefont
  {Eastwood}(1988)}]{Hockney1988}%
  \BibitemOpen
  \bibfield  {author} {\bibinfo {author} {\bibfnamefont {R.}~\bibnamefont
  {Hockney}}\ and\ \bibinfo {author} {\bibfnamefont {J.}~\bibnamefont
  {Eastwood}},\ }\href {http://books.google.fr/books?id=nTOFkmnCQuIC} {\emph
  {\bibinfo {title} {Computer Simulation Using Particles}}}\ (\bibinfo
  {publisher} {Taylor \& Francis},\ \bibinfo {year} {1988})\BibitemShut
  {NoStop}%
\bibitem [{\citenamefont {Birdsall}\ and\ \citenamefont
  {Langdon}(2004)}]{Birdsall2004}%
  \BibitemOpen
  \bibfield  {author} {\bibinfo {author} {\bibfnamefont {C.}~\bibnamefont
  {Birdsall}}\ and\ \bibinfo {author} {\bibfnamefont {A.}~\bibnamefont
  {Langdon}},\ }\href {http://books.google.fr/books?id=S2lqgDTm6a4C} {\emph
  {\bibinfo {title} {Plasma Physics via Computer Simulation}}},\ Series in
  Plasma Physics\ (\bibinfo  {publisher} {Taylor \& Francis},\ \bibinfo {year}
  {2004})\BibitemShut {NoStop}%
\bibitem [{\citenamefont {Esarey}\ \emph {et~al.}(2009)\citenamefont {Esarey},
  \citenamefont {Schroeder},\ and\ \citenamefont {Leemans}}]{EsareyRMP2009}%
  \BibitemOpen
  \bibfield  {author} {\bibinfo {author} {\bibfnamefont {E.}~\bibnamefont
  {Esarey}}, \bibinfo {author} {\bibfnamefont {C.~B.}\ \bibnamefont
  {Schroeder}}, \ and\ \bibinfo {author} {\bibfnamefont {W.~P.}\ \bibnamefont
  {Leemans}},\ }\href {\doibase 10.1103/RevModPhys.81.1229} {\bibfield
  {journal} {\bibinfo  {journal} {Rev. Mod. Phys.}\ }\textbf {\bibinfo {volume}
  {81}},\ \bibinfo {pages} {1229} (\bibinfo {year} {2009})}\BibitemShut
  {NoStop}%
\bibitem [{\citenamefont {Vay}(2007)}]{VayPRL2007}%
  \BibitemOpen
  \bibfield  {author} {\bibinfo {author} {\bibfnamefont {J.-L.}\ \bibnamefont
  {Vay}},\ }\href {\doibase 10.1103/PhysRevLett.98.130405} {\bibfield
  {journal} {\bibinfo  {journal} {Phys. Rev. Lett.}\ }\textbf {\bibinfo
  {volume} {98}},\ \bibinfo {pages} {130405} (\bibinfo {year}
  {2007})}\BibitemShut {NoStop}%
\bibitem [{\citenamefont {Sironi}\ \emph {et~al.}(2013)\citenamefont {Sironi},
  \citenamefont {Spitkovsky},\ and\ \citenamefont {Arons}}]{SironiAJ2013}%
  \BibitemOpen
  \bibfield  {author} {\bibinfo {author} {\bibfnamefont {L.}~\bibnamefont
  {Sironi}}, \bibinfo {author} {\bibfnamefont {A.}~\bibnamefont {Spitkovsky}},
  \ and\ \bibinfo {author} {\bibfnamefont {J.}~\bibnamefont {Arons}},\ }\href
  {http://stacks.iop.org/0004-637X/771/i=1/a=54} {\bibfield  {journal}
  {\bibinfo  {journal} {The Astrophysical Journal}\ }\textbf {\bibinfo {volume}
  {771}},\ \bibinfo {pages} {54} (\bibinfo {year} {2013})}\BibitemShut
  {NoStop}%
\bibitem [{\citenamefont {Godfrey}(1974)}]{GodfreyJCP1974}%
  \BibitemOpen
  \bibfield  {author} {\bibinfo {author} {\bibfnamefont {B.~B.}\ \bibnamefont
  {Godfrey}},\ }\href {\doibase 10.1016/0021-9991(74)90076-X} {\bibfield
  {journal} {\bibinfo  {journal} {Journal of Computational Physics}\ }\textbf
  {\bibinfo {volume} {15}},\ \bibinfo {pages} {504 } (\bibinfo {year}
  {1974})}\BibitemShut {NoStop}%
\bibitem [{\citenamefont {Godfrey}(1975)}]{GodfreyJCP1975}%
  \BibitemOpen
  \bibfield  {author} {\bibinfo {author} {\bibfnamefont {B.~B.}\ \bibnamefont
  {Godfrey}},\ }\href {\doibase 10.1016/0021-9991(75)90116-3} {\bibfield
  {journal} {\bibinfo  {journal} {Journal of Computational Physics}\ }\textbf
  {\bibinfo {volume} {19}},\ \bibinfo {pages} {58 } (\bibinfo {year}
  {1975})}\BibitemShut {NoStop}%
\bibitem [{\citenamefont {Godfrey}\ and\ \citenamefont
  {Vay}(2013)}]{GodfreyJCP2013}%
  \BibitemOpen
  \bibfield  {author} {\bibinfo {author} {\bibfnamefont {B.~B.}\ \bibnamefont
  {Godfrey}}\ and\ \bibinfo {author} {\bibfnamefont {J.-L.}\ \bibnamefont
  {Vay}},\ }\href {\doibase http://dx.doi.org/10.1016/j.jcp.2013.04.006}
  {\bibfield  {journal} {\bibinfo  {journal} {Journal of Computational
  Physics}\ }\textbf {\bibinfo {volume} {248}},\ \bibinfo {pages} {33 }
  (\bibinfo {year} {2013})}\BibitemShut {NoStop}%
\bibitem [{\citenamefont {Xu}\ \emph {et~al.}(2013)\citenamefont {Xu},
  \citenamefont {Yu}, \citenamefont {Martins}, \citenamefont {Tsung},
  \citenamefont {Decyk}, \citenamefont {Vieira}, \citenamefont {Fonseca},
  \citenamefont {Lu}, \citenamefont {Silva},\ and\ \citenamefont
  {Mori}}]{XuCPC2013}%
  \BibitemOpen
  \bibfield  {author} {\bibinfo {author} {\bibfnamefont {X.}~\bibnamefont
  {Xu}}, \bibinfo {author} {\bibfnamefont {P.}~\bibnamefont {Yu}}, \bibinfo
  {author} {\bibfnamefont {S.~F.}\ \bibnamefont {Martins}}, \bibinfo {author}
  {\bibfnamefont {F.~S.}\ \bibnamefont {Tsung}}, \bibinfo {author}
  {\bibfnamefont {V.~K.}\ \bibnamefont {Decyk}}, \bibinfo {author}
  {\bibfnamefont {J.}~\bibnamefont {Vieira}}, \bibinfo {author} {\bibfnamefont
  {R.~A.}\ \bibnamefont {Fonseca}}, \bibinfo {author} {\bibfnamefont
  {W.}~\bibnamefont {Lu}}, \bibinfo {author} {\bibfnamefont {L.~O.}\
  \bibnamefont {Silva}}, \ and\ \bibinfo {author} {\bibfnamefont {W.~B.}\
  \bibnamefont {Mori}},\ }\href {\doibase
  http://dx.doi.org/10.1016/j.cpc.2013.07.003} {\bibfield  {journal} {\bibinfo
  {journal} {Computer Physics Communications}\ }\textbf {\bibinfo {volume}
  {184}},\ \bibinfo {pages} {2503 } (\bibinfo {year} {2013})}\BibitemShut
  {NoStop}%
\bibitem [{\citenamefont {Haber}\ \emph {et~al.}(1973)\citenamefont {Haber},
  \citenamefont {Lee}, \citenamefont {Klein},\ and\ \citenamefont
  {Boris}}]{Haber}%
  \BibitemOpen
  \bibfield  {author} {\bibinfo {author} {\bibfnamefont {I.}~\bibnamefont
  {Haber}}, \bibinfo {author} {\bibfnamefont {R.}~\bibnamefont {Lee}}, \bibinfo
  {author} {\bibfnamefont {H.}~\bibnamefont {Klein}}, \ and\ \bibinfo {author}
  {\bibfnamefont {J.}~\bibnamefont {Boris}},\ }\href@noop {} {\emph {\bibinfo
  {title} {Proc. Sixth Conf. on Num. Sim. Plasmas, Berkeley, CA}}}\ (\bibinfo
  {year} {1973})\BibitemShut {NoStop}%
\bibitem [{\citenamefont {Vay}\ \emph {et~al.}(2010)\citenamefont {Vay},
  \citenamefont {Geddes}, \citenamefont {Benedetti}, \citenamefont {Bruhwiler},
  \citenamefont {Cormier‐Michel}, \citenamefont {Cowan}, \citenamefont
  {Cary},\ and\ \citenamefont {Grote}}]{VayAAC10}%
  \BibitemOpen
  \bibfield  {author} {\bibinfo {author} {\bibfnamefont {J.}~\bibnamefont
  {Vay}}, \bibinfo {author} {\bibfnamefont {C.~G.~R.}\ \bibnamefont {Geddes}},
  \bibinfo {author} {\bibfnamefont {C.}~\bibnamefont {Benedetti}}, \bibinfo
  {author} {\bibfnamefont {D.~L.}\ \bibnamefont {Bruhwiler}}, \bibinfo {author}
  {\bibfnamefont {E.}~\bibnamefont {Cormier‐Michel}}, \bibinfo {author}
  {\bibfnamefont {B.~M.}\ \bibnamefont {Cowan}}, \bibinfo {author}
  {\bibfnamefont {J.~R.}\ \bibnamefont {Cary}}, \ and\ \bibinfo {author}
  {\bibfnamefont {D.~P.}\ \bibnamefont {Grote}},\ }\href {\doibase
  http://dx.doi.org/10.1063/1.3520322} {\bibfield  {journal} {\bibinfo
  {journal} {AIP Conference Proceedings}\ }\textbf {\bibinfo {volume} {1299}},\
  \bibinfo {pages} {244} (\bibinfo {year} {2010})}\BibitemShut {NoStop}%
\bibitem [{\citenamefont {Vay}\ \emph {et~al.}(2011{\natexlab{a}})\citenamefont
  {Vay}, \citenamefont {Geddes}, \citenamefont {Cormier-Michel},\ and\
  \citenamefont {Grote}}]{VayJCP2011}%
  \BibitemOpen
  \bibfield  {author} {\bibinfo {author} {\bibfnamefont {J.-L.}\ \bibnamefont
  {Vay}}, \bibinfo {author} {\bibfnamefont {C.}~\bibnamefont {Geddes}},
  \bibinfo {author} {\bibfnamefont {E.}~\bibnamefont {Cormier-Michel}}, \ and\
  \bibinfo {author} {\bibfnamefont {D.}~\bibnamefont {Grote}},\ }\href
  {\doibase http://dx.doi.org/10.1016/j.jcp.2011.04.003} {\bibfield  {journal}
  {\bibinfo  {journal} {Journal of Computational Physics}\ }\textbf {\bibinfo
  {volume} {230}},\ \bibinfo {pages} {5908 } (\bibinfo {year}
  {2011}{\natexlab{a}})}\BibitemShut {NoStop}%
\bibitem [{\citenamefont {Vay}\ \emph {et~al.}(2011{\natexlab{b}})\citenamefont
  {Vay}, \citenamefont {Geddes}, \citenamefont {Cormier-Michel},\ and\
  \citenamefont {Grote}}]{VayPOPL11}%
  \BibitemOpen
  \bibfield  {author} {\bibinfo {author} {\bibfnamefont {J.-L.}\ \bibnamefont
  {Vay}}, \bibinfo {author} {\bibfnamefont {C.~G.~R.}\ \bibnamefont {Geddes}},
  \bibinfo {author} {\bibfnamefont {E.}~\bibnamefont {Cormier-Michel}}, \ and\
  \bibinfo {author} {\bibfnamefont {D.~P.}\ \bibnamefont {Grote}},\ }\href
  {\doibase http://dx.doi.org/10.1063/1.3559483} {\bibfield  {journal}
  {\bibinfo  {journal} {Physics of Plasmas}\ }\textbf {\bibinfo {volume}
  {18}},\ \bibinfo {eid} {030701} (\bibinfo {year} {2011}{\natexlab{b}}),\
  http://dx.doi.org/10.1063/1.3559483}\BibitemShut {NoStop}%
\bibitem [{\citenamefont {Martins}\ \emph {et~al.}(2010)\citenamefont
  {Martins}, \citenamefont {Fonseca}, \citenamefont {Silva}, \citenamefont
  {Lu},\ and\ \citenamefont {Mori}}]{MartinsCPC10}%
  \BibitemOpen
  \bibfield  {author} {\bibinfo {author} {\bibfnamefont {S.~F.}\ \bibnamefont
  {Martins}}, \bibinfo {author} {\bibfnamefont {R.~A.}\ \bibnamefont
  {Fonseca}}, \bibinfo {author} {\bibfnamefont {L.~O.}\ \bibnamefont {Silva}},
  \bibinfo {author} {\bibfnamefont {W.}~\bibnamefont {Lu}}, \ and\ \bibinfo
  {author} {\bibfnamefont {W.~B.}\ \bibnamefont {Mori}},\ }\href {\doibase
  http://dx.doi.org/10.1016/j.cpc.2009.12.023} {\bibfield  {journal} {\bibinfo
  {journal} {Computer Physics Communications}\ }\textbf {\bibinfo {volume}
  {181}},\ \bibinfo {pages} {869 } (\bibinfo {year} {2010})}\BibitemShut
  {NoStop}%
\bibitem [{\citenamefont {Godfrey}\ and\ \citenamefont
  {Vay}(2014)}]{GodfreyJCP2014b}%
  \BibitemOpen
  \bibfield  {author} {\bibinfo {author} {\bibfnamefont {B.~B.}\ \bibnamefont
  {Godfrey}}\ and\ \bibinfo {author} {\bibfnamefont {J.-L.}\ \bibnamefont
  {Vay}},\ }\href {\doibase 10.1016/j.jcp.2014.02.022} {\bibfield  {journal}
  {\bibinfo  {journal} {Journal of Computational Physics}\ }\textbf {\bibinfo
  {volume} {267}},\ \bibinfo {pages} {1 } (\bibinfo {year} {2014})}\BibitemShut
  {NoStop}%
\bibitem [{\citenamefont {{Godfrey}}(2014)}]{Godfreyarxiv2014}%
  \BibitemOpen
  \bibfield  {author} {\bibinfo {author} {\bibfnamefont {B.~B.}\ \bibnamefont
  {{Godfrey}}},\ }\href@noop {} {\bibfield  {journal} {\bibinfo  {journal}
  {ArXiv e-prints}\ } (\bibinfo {year} {2014})},\ \Eprint
  {http://arxiv.org/abs/1408.1146} {arXiv:1408.1146 [physics.plasm-ph]}
  \BibitemShut {NoStop}%
\bibitem [{\citenamefont {Godfrey}\ and\ \citenamefont
  {Vay}(2015)}]{GodfreyCPC2015}%
  \BibitemOpen
  \bibfield  {author} {\bibinfo {author} {\bibfnamefont {B.~B.}\ \bibnamefont
  {Godfrey}}\ and\ \bibinfo {author} {\bibfnamefont {J.-L.}\ \bibnamefont
  {Vay}},\ }\href {\doibase http://dx.doi.org/10.1016/j.cpc.2015.06.008}
  {\bibfield  {journal} {\bibinfo  {journal} {Computer Physics Communications}\
  ,\ } (\bibinfo {year} {2015})}\BibitemShut {NoStop}%
\bibitem [{\citenamefont {Godfrey}\ \emph
  {et~al.}(2014{\natexlab{a}})\citenamefont {Godfrey}, \citenamefont {Vay},\
  and\ \citenamefont {Haber}}]{GodfreyJCP2014}%
  \BibitemOpen
  \bibfield  {author} {\bibinfo {author} {\bibfnamefont {B.~B.}\ \bibnamefont
  {Godfrey}}, \bibinfo {author} {\bibfnamefont {J.-L.}\ \bibnamefont {Vay}}, \
  and\ \bibinfo {author} {\bibfnamefont {I.}~\bibnamefont {Haber}},\ }\href
  {\doibase http://dx.doi.org/10.1016/j.jcp.2013.10.053} {\bibfield  {journal}
  {\bibinfo  {journal} {Journal of Computational Physics}\ }\textbf {\bibinfo
  {volume} {258}},\ \bibinfo {pages} {689 } (\bibinfo {year}
  {2014}{\natexlab{a}})}\BibitemShut {NoStop}%
\bibitem [{\citenamefont {Godfrey}\ \emph
  {et~al.}(2014{\natexlab{b}})\citenamefont {Godfrey}, \citenamefont {Vay},\
  and\ \citenamefont {Haber}}]{GodfreyIEEE2014}%
  \BibitemOpen
  \bibfield  {author} {\bibinfo {author} {\bibfnamefont {B.}~\bibnamefont
  {Godfrey}}, \bibinfo {author} {\bibfnamefont {J.-L.}\ \bibnamefont {Vay}}, \
  and\ \bibinfo {author} {\bibfnamefont {I.}~\bibnamefont {Haber}},\ }\href
  {\doibase 10.1109/TPS.2014.2310654} {\bibfield  {journal} {\bibinfo
  {journal} {Plasma Science, IEEE Transactions on}\ }\textbf {\bibinfo {volume}
  {42}},\ \bibinfo {pages} {1339} (\bibinfo {year}
  {2014}{\natexlab{b}})}\BibitemShut {NoStop}%
\bibitem [{\citenamefont {Yu}\ \emph {et~al.}(2015{\natexlab{a}})\citenamefont
  {Yu}, \citenamefont {Xu}, \citenamefont {Decyk}, \citenamefont {Fiuza},
  \citenamefont {Vieira}, \citenamefont {Tsung}, \citenamefont {Fonseca},
  \citenamefont {Lu}, \citenamefont {Silva},\ and\ \citenamefont
  {Mori}}]{YuCPC2015}%
  \BibitemOpen
  \bibfield  {author} {\bibinfo {author} {\bibfnamefont {P.}~\bibnamefont
  {Yu}}, \bibinfo {author} {\bibfnamefont {X.}~\bibnamefont {Xu}}, \bibinfo
  {author} {\bibfnamefont {V.~K.}\ \bibnamefont {Decyk}}, \bibinfo {author}
  {\bibfnamefont {F.}~\bibnamefont {Fiuza}}, \bibinfo {author} {\bibfnamefont
  {J.}~\bibnamefont {Vieira}}, \bibinfo {author} {\bibfnamefont {F.~S.}\
  \bibnamefont {Tsung}}, \bibinfo {author} {\bibfnamefont {R.~A.}\ \bibnamefont
  {Fonseca}}, \bibinfo {author} {\bibfnamefont {W.}~\bibnamefont {Lu}},
  \bibinfo {author} {\bibfnamefont {L.~O.}\ \bibnamefont {Silva}}, \ and\
  \bibinfo {author} {\bibfnamefont {W.~B.}\ \bibnamefont {Mori}},\ }\href
  {\doibase http://dx.doi.org/10.1016/j.cpc.2015.02.018} {\bibfield  {journal}
  {\bibinfo  {journal} {Computer Physics Communications}\ }\textbf {\bibinfo
  {volume} {192}},\ \bibinfo {pages} {32 } (\bibinfo {year}
  {2015}{\natexlab{a}})}\BibitemShut {NoStop}%
\bibitem [{\citenamefont {Yu}\ \emph {et~al.}(2015{\natexlab{b}})\citenamefont
  {Yu}, \citenamefont {Xu}, \citenamefont {Tableman}, \citenamefont {Decyk},
  \citenamefont {Tsung}, \citenamefont {Fiuza}, \citenamefont {Davidson},
  \citenamefont {Vieira}, \citenamefont {Fonseca}, \citenamefont {Lu},
  \citenamefont {Silva},\ and\ \citenamefont {Mori}}]{YuCPC2015-Circ}%
  \BibitemOpen
  \bibfield  {author} {\bibinfo {author} {\bibfnamefont {P.}~\bibnamefont
  {Yu}}, \bibinfo {author} {\bibfnamefont {X.}~\bibnamefont {Xu}}, \bibinfo
  {author} {\bibfnamefont {A.}~\bibnamefont {Tableman}}, \bibinfo {author}
  {\bibfnamefont {V.~K.}\ \bibnamefont {Decyk}}, \bibinfo {author}
  {\bibfnamefont {F.~S.}\ \bibnamefont {Tsung}}, \bibinfo {author}
  {\bibfnamefont {F.}~\bibnamefont {Fiuza}}, \bibinfo {author} {\bibfnamefont
  {A.}~\bibnamefont {Davidson}}, \bibinfo {author} {\bibfnamefont
  {J.}~\bibnamefont {Vieira}}, \bibinfo {author} {\bibfnamefont {R.~A.}\
  \bibnamefont {Fonseca}}, \bibinfo {author} {\bibfnamefont {W.}~\bibnamefont
  {Lu}}, \bibinfo {author} {\bibfnamefont {L.~O.}\ \bibnamefont {Silva}}, \
  and\ \bibinfo {author} {\bibfnamefont {W.~B.}\ \bibnamefont {Mori}},\ }\href
  {\doibase 10.1016/j.cpc.2015.08.026} {\bibfield  {journal} {\bibinfo
  {journal} {Computer Physics Communications}\ }\textbf {\bibinfo {volume}
  {197}},\ \bibinfo {pages} {144 } (\bibinfo {year}
  {2015}{\natexlab{b}})}\BibitemShut {NoStop}%
\bibitem [{\citenamefont {{Li}}\ \emph {et~al.}(2016)\citenamefont {{Li}},
  \citenamefont {{Yu}}, \citenamefont {{Xu}}, \citenamefont {{Fiuza}},
  \citenamefont {{Decyk}}, \citenamefont {{Dalichaouch}}, \citenamefont
  {{Davidson}}, \citenamefont {{Tableman}}, \citenamefont {{An}}, \citenamefont
  {{Tsung}}, \citenamefont {{Fonseca}}, \citenamefont {{Lu}},\ and\
  \citenamefont {{Mori}}}]{Yu-arxiv2016}%
  \BibitemOpen
  \bibfield  {author} {\bibinfo {author} {\bibfnamefont {F.}~\bibnamefont
  {{Li}}}, \bibinfo {author} {\bibfnamefont {P.}~\bibnamefont {{Yu}}}, \bibinfo
  {author} {\bibfnamefont {X.}~\bibnamefont {{Xu}}}, \bibinfo {author}
  {\bibfnamefont {F.}~\bibnamefont {{Fiuza}}}, \bibinfo {author} {\bibfnamefont
  {V.~K.}\ \bibnamefont {{Decyk}}}, \bibinfo {author} {\bibfnamefont
  {T.}~\bibnamefont {{Dalichaouch}}}, \bibinfo {author} {\bibfnamefont
  {A.}~\bibnamefont {{Davidson}}}, \bibinfo {author} {\bibfnamefont
  {A.}~\bibnamefont {{Tableman}}}, \bibinfo {author} {\bibfnamefont
  {W.}~\bibnamefont {{An}}}, \bibinfo {author} {\bibfnamefont {F.~S.}\
  \bibnamefont {{Tsung}}}, \bibinfo {author} {\bibfnamefont {R.~A.}\
  \bibnamefont {{Fonseca}}}, \bibinfo {author} {\bibfnamefont {W.}~\bibnamefont
  {{Lu}}}, \ and\ \bibinfo {author} {\bibfnamefont {W.~B.}\ \bibnamefont
  {{Mori}}},\ }\href@noop {} {\bibfield  {journal} {\bibinfo  {journal} {ArXiv
  e-prints}\ } (\bibinfo {year} {2016})},\ \Eprint
  {http://arxiv.org/abs/1605.01496} {arXiv:1605.01496 [physics.comp-ph]}
  \BibitemShut {NoStop}%
\bibitem [{\citenamefont {Lehe}\ \emph
  {et~al.}(2016{\natexlab{a}})\citenamefont {Lehe}, \citenamefont {Kirchen},
  \citenamefont {Godfrey}, \citenamefont {Maier},\ and\ \citenamefont
  {Vay}}]{Lehe2016}%
  \BibitemOpen
  \bibfield  {author} {\bibinfo {author} {\bibfnamefont {R.}~\bibnamefont
  {Lehe}}, \bibinfo {author} {\bibfnamefont {M.}~\bibnamefont {Kirchen}},
  \bibinfo {author} {\bibfnamefont {B.~B.}\ \bibnamefont {Godfrey}}, \bibinfo
  {author} {\bibfnamefont {A.~R.}\ \bibnamefont {Maier}}, \ and\ \bibinfo
  {author} {\bibfnamefont {J.-L.}\ \bibnamefont {Vay}},\ }\href@noop {}
  {\bibfield  {journal} {\bibinfo  {journal} {to be submitted}\ } (\bibinfo
  {year} {2016}{\natexlab{a}})}\BibitemShut {NoStop}%
\bibitem [{\citenamefont {Vay}\ \emph {et~al.}(2012)\citenamefont {Vay},
  \citenamefont {Grote}, \citenamefont {Cohen},\ and\ \citenamefont
  {Friedman}}]{Warpref}%
  \BibitemOpen
  \bibfield  {author} {\bibinfo {author} {\bibfnamefont {J.-L.}\ \bibnamefont
  {Vay}}, \bibinfo {author} {\bibfnamefont {D.~P.}\ \bibnamefont {Grote}},
  \bibinfo {author} {\bibfnamefont {R.~H.}\ \bibnamefont {Cohen}}, \ and\
  \bibinfo {author} {\bibfnamefont {A.}~\bibnamefont {Friedman}},\ }\href
  {http://stacks.iop.org/1749-4699/5/i=1/a=014019} {\bibfield  {journal}
  {\bibinfo  {journal} {Computational Science \& Discovery}\ }\textbf {\bibinfo
  {volume} {5}},\ \bibinfo {pages} {014019} (\bibinfo {year}
  {2012})}\BibitemShut {NoStop}%
\bibitem [{\citenamefont {Lifschitz}\ \emph {et~al.}(2009)\citenamefont
  {Lifschitz}, \citenamefont {Davoine}, \citenamefont {Lefebvre}, \citenamefont
  {Faure}, \citenamefont {Rechatin},\ and\ \citenamefont {Malka}}]{Lifschitz}%
  \BibitemOpen
  \bibfield  {author} {\bibinfo {author} {\bibfnamefont {A.~F.}\ \bibnamefont
  {Lifschitz}}, \bibinfo {author} {\bibfnamefont {X.}~\bibnamefont {Davoine}},
  \bibinfo {author} {\bibfnamefont {E.}~\bibnamefont {Lefebvre}}, \bibinfo
  {author} {\bibfnamefont {J.}~\bibnamefont {Faure}}, \bibinfo {author}
  {\bibfnamefont {C.}~\bibnamefont {Rechatin}}, \ and\ \bibinfo {author}
  {\bibfnamefont {V.}~\bibnamefont {Malka}},\ }\href {\doibase
  10.1016/j.jcp.2008.11.017} {\bibfield  {journal} {\bibinfo  {journal} {J.
  Comput. Phys.}\ }\textbf {\bibinfo {volume} {228}},\ \bibinfo {pages} {1803}
  (\bibinfo {year} {2009})}\BibitemShut {NoStop}%
\bibitem [{\citenamefont {Lehe}\ \emph
  {et~al.}(2016{\natexlab{b}})\citenamefont {Lehe}, \citenamefont {Kirchen},
  \citenamefont {Andriyash}, \citenamefont {Godfrey},\ and\ \citenamefont
  {Vay}}]{LeheCPC2016}%
  \BibitemOpen
  \bibfield  {author} {\bibinfo {author} {\bibfnamefont {R.}~\bibnamefont
  {Lehe}}, \bibinfo {author} {\bibfnamefont {M.}~\bibnamefont {Kirchen}},
  \bibinfo {author} {\bibfnamefont {I.~A.}\ \bibnamefont {Andriyash}}, \bibinfo
  {author} {\bibfnamefont {B.~B.}\ \bibnamefont {Godfrey}}, \ and\ \bibinfo
  {author} {\bibfnamefont {J.-L.}\ \bibnamefont {Vay}},\ }\href {\doibase
  http://dx.doi.org/10.1016/j.cpc.2016.02.007} {\bibfield  {journal} {\bibinfo
  {journal} {Computer Physics Communications}\ }\textbf {\bibinfo {volume}
  {203}},\ \bibinfo {pages} {66 } (\bibinfo {year}
  {2016}{\natexlab{b}})}\BibitemShut {NoStop}%
\bibitem [{\citenamefont {Lehe}\ \emph {et~al.}(2013)\citenamefont {Lehe},
  \citenamefont {Lifschitz}, \citenamefont {Thaury}, \citenamefont {Malka},\
  and\ \citenamefont {Davoine}}]{LehePRSTAB2013}%
  \BibitemOpen
  \bibfield  {author} {\bibinfo {author} {\bibfnamefont {R.}~\bibnamefont
  {Lehe}}, \bibinfo {author} {\bibfnamefont {A.}~\bibnamefont {Lifschitz}},
  \bibinfo {author} {\bibfnamefont {C.}~\bibnamefont {Thaury}}, \bibinfo
  {author} {\bibfnamefont {V.}~\bibnamefont {Malka}}, \ and\ \bibinfo {author}
  {\bibfnamefont {X.}~\bibnamefont {Davoine}},\ }\href {\doibase
  10.1103/PhysRevSTAB.16.021301} {\bibfield  {journal} {\bibinfo  {journal}
  {Phys. Rev. ST Accel. Beams}\ }\textbf {\bibinfo {volume} {16}},\ \bibinfo
  {pages} {021301} (\bibinfo {year} {2013})}\BibitemShut {NoStop}%
\bibitem [{\citenamefont {Vay}\ \emph {et~al.}(2013)\citenamefont {Vay},
  \citenamefont {Haber},\ and\ \citenamefont {Godfrey}}]{VayJCP2013}%
  \BibitemOpen
  \bibfield  {author} {\bibinfo {author} {\bibfnamefont {J.-L.}\ \bibnamefont
  {Vay}}, \bibinfo {author} {\bibfnamefont {I.}~\bibnamefont {Haber}}, \ and\
  \bibinfo {author} {\bibfnamefont {B.~B.}\ \bibnamefont {Godfrey}},\ }\href
  {\doibase http://dx.doi.org/10.1016/j.jcp.2013.03.010} {\bibfield  {journal}
  {\bibinfo  {journal} {Journal of Computational Physics}\ }\textbf {\bibinfo
  {volume} {243}},\ \bibinfo {pages} {260 } (\bibinfo {year}
  {2013})}\BibitemShut {NoStop}%
\bibitem [{\citenamefont {Vincenti}\ and\ \citenamefont
  {Vay}(2016)}]{Vincenti2015}%
  \BibitemOpen
  \bibfield  {author} {\bibinfo {author} {\bibfnamefont {H.}~\bibnamefont
  {Vincenti}}\ and\ \bibinfo {author} {\bibfnamefont {J.-L.}\ \bibnamefont
  {Vay}},\ }\href {\doibase http://dx.doi.org/10.1016/j.cpc.2015.11.009}
  {\bibfield  {journal} {\bibinfo  {journal} {Computer Physics Communications}\
  }\textbf {\bibinfo {volume} {200}},\ \bibinfo {pages} {147 } (\bibinfo {year}
  {2016})}\BibitemShut {NoStop}%
\end{thebibliography}
\end{document}